# Self-Organization and Finite Velocity Of Transmitting Substance and Energy through Space-Time

**Maria K. Koleva**
*Institute of Catalysis, Bulgarian Academy of Sciences, 1113 Sofia, Bulgaria*
*e-mail: mkoleva@bas.bg*

The idea that the velocity of transmitting substance/energy trough space-time is to be bounded is a fundamental concept in the science. To the most surprise, it turns out that it is not always met. We demonstrate that the existing approaches to the self-organization, another major concept in the science, let the velocity of transmitting substance to be arbitrary. Further we prove that only the boundedness of the velocity is not enough to ensure the self-organization. That is why we develop radically novel approach to the macroscopic evolution that not only reconciles the self-organization and the velocity ansatz but in addition gives physically credible basis to phenomena like Feigenbaum cascade and fluctuation-assisted bifurcations.

**Pacs**: 05.65.+b; 05.40.-a

## Introduction

The self-organization is one of the most advanced and fascinating hypothesis about the behavior of the extended many-body systems with short-range interactions. It asserts that the emergent properties on macroscopic scale, obtained as a result of self-organization, are insensitive to the microscopic dynamics and are independent on the individual elements and components. The mild conditions of the derivation of the reaction-diffusion equations (hyperbolic partial differential equations) that serves as mathematical tool for its description support the ubiquity of its application: the idea of self-organization is applied to wide spectrum of systems as diverse as chemical systems, optical lattices, social systems, population dynamics etc. Indeed, these equations are founded on the postulate that the diffusion does not create or annihilate substance which provides the conservation of substance flow in space-time. However, the rigorous derivation of the reaction-diffusion equations requires explicit demonstration that the conditions at which the microscopic dynamics does not contribute to the emergent properties are as credible as the idea of flux conservation. In particular, one such condition is that the local fluctuations are to be automatically damped. Moreover, it requires that damping happens at time scale much smaller than the specific scale on which the state variables change significantly. This is a crucial point, since the separation of the time scales to the fast and slow ones provides spatio-temporal continuality of the intensive state variables such as concentrations, temperature etc. so that the flux of substance to be expressed in terms of those variables. In turn, this ensures the target independence of the emergent properties from the microscopic dynamics. But does the separation of the time scales is grounded on the same plausible basis as the substance conservation? Our first aim is to demonstrate that, to the most surprise, it is not. Next we provide evidence that the basic existing models approach the separation of the time scales letting the velocity of transmitting substance trough space to be arbitrary. Note, that arbitrary means that the velocity can be greater than the speed of light! On the other hand, as we will demonstrate further, the boundedness of the velocity alone is not sufficient to make the self-organization available because it participates in exponential amplification of the local fluctuations. That amplification is a novel effect, considered for the first time in &1, that is result of the interplay between the boundedness of the velocity and the lack of correlations among local fluctuations due to short-rang interactions. In turn, the amplification of the local fluctuations assisted by the lack of correlation among them results in triggering of local destabilization that would yield system breakdown if not eliminated. Therefore, as it will become evident later, the only way to the elimination of the amplification goes via non-local coupling of the local fluctuations that makes their further response coherent and comprises physical agent that spreads it throughout the entire system. The fundamental problem is that the target coherence should persist at every value of the external constraints imposed on the system. This requirement makes the coherence fundamentally different from phase transitions that take place only at specific values of the external constraints. We have already established [1] that a successful coherence mechanism operates only on the grounds of radically novel viewpoint on the interactions. In &2-&3 we shall demonstrate that the interplay between the boundedness ansatz and the coherence saves the hypothesis of self-organization at the expense of introducing radically novel features. But prior to that we shall illustrate that short-range interactions gives rise to permanent destabilization that is locally amplified even when the velocity ansatz holds. To elucidate better the problem let us first make apparent how the self-organization appears when the velocity of transmitting substance is let to be arbitrary. For this purpose let us follow Gardiner [2] in the derivation of the diffusion equation:

Let $f(x,t)$ be the number of entities per unit volume. We compute the distribution of entities at the time $t+\tau$ from the distribution at time $t$. Let the function $\phi(\Delta)$ be the frequency of the jumps to distance $\Delta$ back and forth a given point. Then, the flux conservation requires that the number of entities which at time $t+\tau$ are found between two planes perpendicular to the $x-$ axis and passing through points $x$ and $x+dx$. One obtains:

$$f(x, t+\tau)dx = dx \int_{-\infty}^{+\infty} f(x+\Delta)\phi(\Delta)d\Delta \qquad (1)$$

*Under the supposition that $\tau$ is very small, we can set:*

$$f(x, t+\tau) = f(x,t) + \tau \frac{\partial f}{\partial t} \qquad (2)$$

*Furthermore, we develop $f(x+\Delta, t)$ in powers of $\Delta$:*

$$f(x+\Delta, t) = f(x,t) + \Delta \frac{\partial f(x,t)}{\partial x} + \frac{\Delta^2}{2!} \frac{\partial^2 (x,t)}{\partial x^2} + \ldots \qquad (3)$$

*We can use this series under the integral:*

$$f + \frac{\partial f}{\partial t}\tau = f \int_{-\infty}^{+\infty} \phi(\Delta)d\Delta + \frac{\partial f(x,t)}{\partial x} \int_{-\infty}^{+\infty} \Delta \phi(\Delta)d\Delta + \frac{\partial^2 f(x,t)}{\partial x^2} \int_{-\infty}^{+\infty} \frac{\Delta^2}{2} \phi(\Delta)d\Delta + \ldots \qquad (4)$$

*Because $\phi(x) = \phi(-x)$, the second, fourth etc. terms on the right-hand side vanish, while out of the 1st, 3rd, 5th etc., terms, each small compared with the previous. We obtain from this equation, by taking into account consideration:*

$$\int_{-\infty}^{+\infty} \phi(\Delta)d\Delta = 1 \qquad (5)$$

*and setting*

$$\frac{1}{\tau} \int_{-\infty}^{+\infty} \frac{\Delta^2}{2} \phi(\Delta)d\Delta = D \qquad (6)$$

*Keeping only the 1st and third term of the right-hand side,*

$$\frac{\partial f}{\partial t} = D \frac{\partial^2 f}{\partial x^2} \qquad (7)$$

In result, the microscopic derivation yields the flux conservation. Moreover, the dependence on the microscopic dynamics is encapsulated in a single parameter $D$ that comprises asymptotic statistical property, namely the variance of the jump length. However, the circle is not yet closed because the above derivation comprises additional supposition that has not been considered. Indeed, let us have a closer look on the integration in (1) and (4). The philosophy of (4) asserts that in a small but otherwise arbitrary time interval $\tau$ the flux that crosses the boundary between $x$ and $x+dx$ is the average between jumps back and forth the layer. Yet, the range of integration of the jump length $(-\infty, +\infty)$ implicates independence of the jump length $\Delta$ and the duration of the time interval $\tau$ from one another. In turn, that independence renders the microscopic dynamics to be averaged out only on times scales larger than certain one. However, this brings about an apparent contradiction: on the one hand, the credibility of the expansions in (1)-(4) implies existence of certain time scale $t_{fast}$ over which the microscopic dynamics is averaged out. Therefore, the value of $\tau$ is to be set larger than $t_{fast}$; on the other hand the notion of the diffusion coefficient $D$ in (7) requires $\tau$ to be smaller than $t_{fast}$. Moreover, the existence of $D$ is possible if and only if $\Delta$ and $\tau$ are related; otherwise, the lack of relation between $\Delta$ and $\tau$ renders $D$ $\tau-$ dependent provided the variance of the jump length exists. Thus, one and the same $\tau$ participates to the same equation in two confronting each other roles.

In addition, the independence of the length of a jump $\Delta$ and its duration $\tau$ suffers severe physical disadvantage: it renders the velocity $\frac{\Delta}{\tau}$ of transmitting substance trough space to be arbitrary; in particular it can become even greater than the speed of the light!

Note, that the lack of relation between $\Delta$ and $\tau$ is plausible only under the condition of perfect stirring which asserts that stirring is so intensive that every entity can be found with equal probability anywhere in the system. However, the stirring is not available for a number of systems such as interfaces, optical lattices, social systems, population dynamics etc. The other interpretation of the perfect stirring, namely that the system has already arrived at local thermodynamical equilibrium, does not help reconciliation because it does *not* point out the route through which the microscopic dynamics ensures the thermodynamical equilibrium.

So, in order to put the hypothesis about self-organization on stable ground, the starting point must be the postulate that whatever the system is the velocity of transmitting substance and/or energy is to be bounded (velocity ansatz).

## 1. Amplification of the Local Fluctuations

The question that immediately arises is whether the requirement about relation between $\Delta$ and $\tau$, i.e. velocity ansatz, is enough to save the idea of self-organization. The goal of the present section is to demonstrate that: (i) the dependence between $\Delta$ and $\tau$ gives rise to local fluctuations; (ii) due to short-range interactions, the lack of correlations among them makes possible their amplification. In result, the latter not only violates the idea of self-organization but gives rise to permanent destabilization of the system that rapidly yields its breakdown if not eliminated. In the next section we will demonstrate that the elimination of the destabilization is possible only under radically novel viewpoint on a number of issues.

Let us first see how the diffusion equation is modified under the postulate about boundedness of the velocity. Obviously, it makes $\Delta$ function of $\tau : \Delta = \Delta(\tau)$. Then, (1) is modified as follows:

$$f(x, t+\tau)dx = dx \int_{-\zeta(\tau)}^{+\zeta(\tau)} f(x+\Delta(\tau))\phi(\Delta(\tau))d\tau \tag{8}$$

where $\zeta(\tau)$ is the maximum possible jump length at given $\tau$. Accordingly, eq.(4) becomes:

$$f + \frac{\partial f}{\partial t}\tau = f\int_{-\zeta(\tau)}^{+\zeta(\tau)}\phi(\Delta)d\Delta + \frac{\partial f(x,t)}{\partial x}\int_{-\zeta(\tau)}^{+\zeta(\tau)}\Delta(\tau)\phi(\Delta(\tau))d\tau + \frac{\partial^2 f(x,t)}{\partial x^2}\int_{-\zeta(\tau)}^{+\zeta(\tau)}\frac{\Delta^2(\tau)}{2}\phi(\Delta(\tau))d\tau + \ldots \tag{9}$$

The fundamental difference between (4) and (9) is that the averaging in the right-hand side of the integrals in (4) specifies only the range of the jump length leaving the time interval arbitrary while in (9) the averaging explicitly involves the dependence between the jump length and its duration. In result, the averaging over a finite time interval yields:

$$\int_{-\zeta(\tau)}^{+\zeta(\tau)}\Delta(\tau)\phi(\Delta(\tau))d\tau = \beta(x,t,\zeta(\tau)) \tag{10}$$

where $\beta(x,t,\zeta(\tau))$ is non-zero and changes its sign from positive to negative depending on the space-time location of the window $\zeta(\tau)$. To compare, the averaging over infinite interval makes $\beta(x,t,\zeta(\tau)) \equiv 0$ at every spatio-temporal point $(x,t)$:

$$\beta(x,t,\zeta(\tau)) = \int_{-\infty}^{+\infty}\Delta\phi(\Delta)d\Delta \equiv 0 \tag{11}$$

Accordingly:

$$\int_{-\zeta(\tau)}^{+\zeta(\tau)}\frac{\Delta^2(\tau)}{2}\phi(\Delta(\tau))d\tau = D(x,t,\zeta(\tau)) \tag{12}$$

where the value of $D(x,t,\zeta(\tau))$ varies with the size and location of the windows. In result, the equation for $f(x,t)$ becomes:

$$\frac{\partial f(x,t,\zeta(\tau))}{\partial t} = \beta(x,t,\zeta(\tau))\frac{\partial f(x,t,\zeta(\tau))}{\partial x} + D(x,t,\zeta(\tau))\frac{\partial^2 f(x,t,\zeta(\tau))}{\partial x^2} \tag{13}$$

The fundamental difference between (7) and (13) is that in (13) the window parameters $\zeta$ and $\tau$ are entangled with the spatio-temporal variables $x$ and $t$ while in (7) the window parameters do not participate at all. However, the entanglement of the window parameters and the spatio-temporal variables strongly interferes with the idea of self-organization that the emergent properties on macroscopic scale are to be independent of the microscopic dynamics. Besides, it opens the door to the following option: since the lack of correlations among local fluctuations renders $\beta(x,t,\zeta(\tau))$ and $D(x,t,\zeta(\tau))$ to be irregular functions, is it possible that their stochasticity brings about amplification of the local fluctuations that in turn triggers local destabilization followed by rapidly developed global one. In result we face a fundamental problem: on the one hand, in order to save the idea of self-organization we must save the supposition about the existence of two time scales $t_{slow}$ and $t_{fast}$ so that:

$$f(x,t,\zeta(\tau)) = f_{slow}(x,t) + f_{fast}(x,t,\zeta(\tau)) \tag{14}$$

(ii) on the other hand, the supposition about separability of the times scales is justified only by automatic damping of the local fluctuations. Next, we will demonstrate that the fulfillment of both requirements is impossible when the local fluctuations are non-correlated. We proceed through the opposite, namely we suppose *apriori* that the separation of the time scales holds. Our task is to prove that it does not prevent amplification of the local fluctuations.

The separation of the time scales implies that for the time scales $\Delta t$ such that $t_{slow} \gg \Delta t \gg t_{fast}$, the following averaging yields:

$$\int_0^{\Delta t} f_{fast}(x,t,\zeta(\tau))d\tau = 0$$

$$\int_0^{\Delta t} \frac{\partial f_{fast}(x,t,\zeta(\tau))}{\partial x}d\tau = 0 \tag{15}$$

$$\int_0^{\Delta t} \frac{\partial^2 f_{fast}(x,t,\zeta(\tau))}{\partial x^2}d\tau = 0$$

The condition (15) actually imposes uniform convergence of $f_{fast}(x,t,\zeta(\tau))$ and its first and second derivative to $f_{slow}(x,t)$ and its corresponding derivatives. Note that (15) is justified whenever the local fluctuations are damped on the time scales smaller than $t_{fast}$: if so, the distance between $f_{fast}(x,t,\zeta(\tau))$ and $f_{slow}(x,t)$ is always finite which in turn verifies both the separation in (14) and the uniform convergence set by (15). As an immediate consequence of (14) and (15), the dependence of $f(x,t,\zeta(\tau))$ on the microscopic dynamics, i.e. on the window parameters $(\zeta,\tau)$ and their relation, is limited to the time intervals smaller than $t_{fast}$. Now we come to the question at what conditions on $\beta(x,t,\zeta(\tau))$ and $D(x,t,\zeta(\tau))$ (14)-(15) hold. Let us now substitute $f(x,t,\zeta(\tau))$ from (14) in (13). It is obvious that a necessary condition for (15) to hold is that $\beta(x,t,\zeta(\tau))$ and $D(x,t,\zeta(\tau))$ additively decompose to slow and fast parts:

$$\beta(x,t,\zeta(\tau)) = \beta_{slow}(x,t) + \beta_{fast}(x,t,\zeta(\tau)) \tag{16}$$
$$D(x,t,\zeta(\tau)) = D_{slow}(x,t) + D_{fast}(x,t,\zeta(\tau)) \tag{17}$$

In general, both $\beta_{fast}(x,t,\zeta(\tau))$ and $D_{fast}(x,t,\zeta(\tau))$ must be irregular functions that fulfill the following condition: their average over window of any length greater than $t_{fast}$ uniformly converges to $\beta_{slow}(x,t)$ and $D_{slow}(x,t)$ correspondingly. This requirement opens the door for $D_{fast}(x,t,\zeta(\tau))$ to change sign. We shall demonstrate that the negative values of the diffusion coefficient implement the amplification of the local fluctuations. Furthermore, the condition that the walk is symmetric renders $\beta_{slow}(x,t) \equiv 0$. Then, eq.(13) becomes:

$$\frac{\partial f_{slow}(x,t)}{\partial t} + \frac{\partial f_{fast}(x,t,\zeta(\tau))}{\partial t} = \beta_{fast}(x,t,\zeta(\tau))\frac{\partial f_{slow}(x,t)}{\partial x} + \beta_{fast}(x,t,\zeta(\tau))\frac{\partial f_{fast}(x,t,\zeta(\tau))}{\partial x} + \\ + (D_{slow}(x,t) + D_{fast}(x,t,\zeta(\tau)))\left(\frac{\partial^2 f_{slow}(x,t)}{\partial x^2} + \frac{\partial^2 f_{fast}(x,t,\zeta(\tau))}{\partial x^2}\right)$$
(18)

Let us now average over an interval $\Delta t$ such that $t_{slow} >> \Delta t >> t_{fast}$. Then, on the condition for uniform convergence, (18) decouples as follows:

$$\frac{\partial f_{slow}(x,t)}{\partial t} = D_{slow}(x,t)\frac{\partial^2 f(x,t)}{\partial x^2}$$
(19)

and

$$\frac{\partial f_{fast}(x,t,\zeta(\tau))}{\partial t} = \beta_{fast}(x,t,\zeta(\tau))\frac{\partial f_{fast}(x,t,\zeta(\tau))}{\partial x} + D_{fast}(x,t,\zeta(\tau))\frac{\partial^2 f_{fast}(x,t,\zeta(\tau))}{\partial x^2}$$
(20)

The terms that contain one slow and one fast multiplier becomes zero after averaging over $\Delta t$ because the corresponding slow term remains constant over that time interval $\Delta t << t_{slow}$. For example:

$$\left\langle D_{slow}(x,t)\frac{\partial^2 f_{fast}(x,t,\zeta(\tau))}{\partial x^2}\right\rangle = D_{slow}(x,t)\left\langle \frac{\partial^2 f_{fast}(x,t,\zeta(\tau))}{\partial x^2}\right\rangle = 0$$
(21)

However, the averaging in (20) commuts with the differentiation both in time and space only if $f_{fast}(x,t,\zeta(\tau))$ is always finite; the latter is provided only by automatic damping of the local fluctuations. On the contrary, the alternative of amplification opens the door to unlimited increasing of $f_{fast}(x,t,\zeta(\tau))$. In order to study that option we have to examine the properties of (20) over time scales smaller than $t_{fast}$. For this purpose let us now present (20) in the following slightly modified form:

$$\frac{\partial f_{fast}(x,t,\zeta(\tau))}{\partial t} = \frac{\partial}{\partial x}\left(\tilde{D}_{fast}(x,t,\zeta(\tau))\frac{\partial f_{fast}(x,t,\zeta(\tau))}{\partial x}\right)$$
(22)

where $\tilde{D}_{fast}(x,t,\zeta(\tau))$ is such that:

$$\frac{\partial \tilde{D}_{fast}(x,t,\zeta(\tau))}{\partial x} = \beta(x,t,\zeta(\tau))$$
(23)

Since both $D_{fast}(x,t,\zeta(\tau))$ and $\beta(x,t,\zeta(\tau))$ are irregular functions with identical statistical properties, $\tilde{D}_{fast}(x,t,\zeta(\tau))$ is again an irregular function that share the same statistical properties. Actually, the value and sign of $\tilde{\tilde{D}}_{fast}(x,t,\zeta(\tau))$ depend mainly is on the window parameters $\zeta$ and $\tau$; the dependence on spatio-temporal variables $(x,t)$ serves to point out explicitly the lack of correlations among their values at closest spatio-temporal locations. This lets us to regard $\tilde{D}_{fast}(x,t,\zeta(\tau))$ as function of the window parameters only. In turn, this consideration let us approximate (22) as follows:

$$\frac{\partial f_{fast}(x,t,\zeta(\tau))}{\partial t} = \tilde{D}_{fast}(x,t,\zeta(\tau))\frac{\partial^2 f_{fast}(x,t,\zeta(\tau))}{\partial x^2} \qquad (24)$$

Note that $\tilde{D}_{fast}(x,t,\zeta(\tau))$ is explicitly grounded on the velocity ansatz! This sets the fundamental difference between the present approach and the Langvin one because the latter uses Wiener process for modeling the lack of correlations among local fluctuations. The problem is that the unboundedness of the increments of the Wiener process allows transmitting of arbitrary amount of substance with arbitrary velocity. On the contrary, our goal is to derive the emergent hypothesis on the grounds of the postulate about the finite velocity of transmitting substance/energy.

The consideration that $\tilde{D}_{fast}(x,t,\zeta(\tau))$ is set constant equal to its current value opens the door for explicit establishing the solution. The velocity ansatz bounds us to look for the solution in the form of a single variable $\lambda$ that couples $x$ and $t$ so that the velocity to be bounded. Then $\lambda = \frac{x}{t^{\alpha(x/t)}}$ where we set $\alpha\left(\frac{x}{t}\right)$ function because the diffusion coefficient $\tilde{D}_{fast}(x,t,\zeta(\tau))$ varies in the space-time. Therefore, the major term in the solution of (24) reads:

$$f_{fast}(x,t,\zeta(\tau)) \propto \exp\left(-\frac{x^2}{\tilde{D}(x,t,\zeta(\tau))t}\right) \qquad (25)$$

Since the value and sign of $\tilde{D}_{fast}(x,t,\zeta(\tau))$ varies form on location to the next, $f_{fast}(x,t,\zeta(\tau))$ is also an irregular function that grows up exponentially when $\tilde{D}_{fast}(x,t,\zeta(\tau))$ is negative. Thus, the corresponding local fluctuations are amplified. Moreover, though the rate of the obtained amplification depends only on the local window parameters, it can happen at every instant and everywhere in the system. In turn, it gives rise to and sustains permanent destabilization of the system.

It should be stressed that the reaction-diffusion coupling is not able to prevent the destabilization since the reaction interactions are local events the rate of which is proportional to the local concentrations of the reactants. So, the local fluctuations are straightforwardly mapped onto the reaction rates. This consideration is easy to be traced in the formal solution of the corresponding reaction-diffusion equations that looks like as follows:

$$\frac{\partial f_{fast}(x,t,\zeta(\tau))}{\partial t} = \kappa A(f_{fast}(x,t,\zeta(\tau))) - \gamma R(f_{fast}(x,t,\zeta(\tau))) + \tilde{D}_{fast}(x,t,\zeta(\tau))\frac{\partial^2 f_{fast}(x,t,\zeta(\tau))}{\partial x^2} \qquad (26)$$

where $\kappa$ and $\gamma$ are the control parameters; $A(f_{fast}(x,t,\zeta(\tau)))$ is the rate by which the substance enters the system; $R(f_{fast}(x,t,\zeta(\tau)))$ is the reaction rate. Since both (24) and (26) are hyperbolic partial differential equation that share the same Green function whose major term is (25), the reaction-diffusion coupling indeed does not help preventing the amplification of the local fluctuations.

In turn, a successful elimination of the obtained destabilization is possible only if there is a mechanism that drives the local fluctuations to behave coherently. Prior to any further consideration, it should be pointed out that the target coherence is fundamentally different from the phase transitions since they take place only at certain values of the external constraints while the coherence should operate at every value of the external constraints.

Moreover, since without doubt the coherence should be grounded on the velocity ansatz, the only route to elimination of the destabilization goes through looking for long-range interactions that serve as physical agent that provides the target coherence on the condition that this process meets it. Note that the prevention of the amplification cannot be accomplished by means of collisions only. Indeed, they yield local evening of the current states of colliding entities. However, due to the lack of correlations, that evening gives rise both to amplification of the interactions in one location and to their damping in other one. Thus, to the most surprise, the collisions contribute to the amplification of the local fluctuations! Therefore, the only way to the elimination of the amplification goes via non-local coupling that results in coherent response and comprises physical agent that spreads it throughout the entire system.

## 2. Adaptation and Coherence

One of the greatest challenges to the coherence is imposed by its ubiquity: since it is supposed to operate in a wide spectrum of systems, the physical interactions that serve as operational agent of its realization should be insensitive to the chemical identity of the entities that constitute the system. Thus, we face the fundamental problem: how to reconcile the chemical identity and the universality of those interactions. We start revealing of the puzzle by recalling how the chemical identity is incorporated in the microscopic dynamics. It is generally accepted that the bearer of identity of an entity is its Hamiltonian $\hat{H}_0$. Further, the interactions of that entity with the environment are considered under the concept that the physical interactions do not change the identity. On the grounds of that postulate, the interaction with the environment additively decomposes to:

$$\hat{H}_{tot} = \hat{H}_0 + \lambda \sum_{i=1}^{N} \hat{V}_i(t) \tag{27}$$

where $\hat{H}_{tot}$ is the total Hamiltonian; $\hat{V}_i(t)$ is the interaction of the given entity with the $i-th$ one; $\lambda$ is a parameter that measures the intensity of the interaction; $N$ is number of the interactions that contribute to the value of $\lambda$. Note that the idea that the interactions do not change the identity renders $\hat{H}_0$ time-independent. However, the success of the additive decomposition expressed by (27) is limited to the case of weak coupling to the environment, i.e. it is credible only when $\lambda << 1$. Obviously, $\lambda$ can be kept permanently small only when the interactions are short-ranged since the latter implicates that $N$ is limited to the immediate neighborhood of the given entity.

However, the obtained in the previous section amplification of the local fluctuations and the lack of correlations among them let the interactions to be intensified to arbitrary large values. In turn, if not suppressed, the unlimited intensification certainly causes the entity breakdown. Therefore, the long-term stability imposes the following general constraint to the relation entity-environment: in order to prevent breakdown, there should be a mechanism that permanently suppress the unlimited intensification of the interactions. We assert that this mechanism is coherence. Yet, as we shall see later, its successful operation it is grounded on the following idea: on reaching critical intensity the interactions $\hat{H}_0$ changes non-perturbatively so that to open the door for weakening of the total interaction by means of "switching off" some of the contributing interactions. For example, an appropriate change of the symmetry turns the magnetic moment to zero which in turn eliminates the corresponding magnetic interaction. These considerations make reasonable to assume that the identity whose bearer is $\hat{H}_0$ is a property that modifies upon reaching critical intensity of the interaction. In turn, the modification helps the entity to "adapt" to permanently changing environment by means of local switching on and off interactions so that to keep the total interaction energy permanently bounded, i.e. below certain thresholds. The mathematical implement of the adaptation is the operation of coarse-graining that replaces the linear superposition (27); unlike the linear superposition which allows unlimited intensification, the coarse-graining is a non-homogeneous non-linear operation that preserves boundedness. The onset of its non-homogeneity and non-linearity is the switching on and off the interactions which is equivalent to local amplification/damping of the total interaction.

However, it seems that the above idea of adaptation has serious flaw: on the one hand, the adaptation is achieved at the expense of lost identity; on the other hand the spectroscopy and chemistry confirm in a very categorical way the undisturbed identity in a number of processes. The compromise between them points that the way out is to suppose decomposition of $\hat{H}_0$ as follows:

$$\hat{H}_0 = \hat{H}_{rigid} \oplus \hat{H}_{flexible} \tag{28}$$

where $\hat{H}_{rigid}$ is assumed to save the identity and $\hat{H}_{flexible}$ is assumed to modify its identity under the influence of the environment. Our next task is to demonstrate that (28) is not only a formal compromise but has strong physical origin as well. The pivotal clue is the question whether there is any additional factor that contribute to the adaptation; and if so, does its action differs for different parts of $\hat{H}_0$. The answer is affirmative – there is such factor and it is related to the dualism wave-particle. Indeed, one of the outcomes of that dualism is that the structure of a particle becomes "smoothed out" at distances up to its de Broglie wave-length. Thus, when the latter is of the order of the size of the interaction region, the blockade of its symmetries renders the particle to behave as structureless point-mass. In result, the inactivation of the symmetries acts towards weakening of the total interaction independently from the adaptation. However, the symmetry blockade is pronounced only for the states of low energies, i.e. for the states whose de Broglie wave-length is large enough. In result, the highly excited states loose their symmetries at intensity of interactions smaller than critical for triggering the adaptation one. In result, the very highly excited states loose all their symmetries while the states of lowest energies are still subject to linear superposition. In turn this justifies the decomposition (28).

Yet, the idea of adaptation strongly challenges the notion of Hamiltonian because it renders $\hat{H}_{flexible}$ time-dependent. Therefore, is it possible to expect any binding associated with it? It turns out that the coarse-graining provides tricky interplay between binding and scattering: the binding in a given location is temporary and is followed by transition to another binding region; after a while a new transition to other binding region takes place etc. The properties of that behavior have been systematically studied in Chapter 5 of [1]. It has been demonstrated that the spectrum of each $\hat{H}_{flexible}$ is characterized by a single parameter, energy. Moreover, the level spacing of all $\hat{H}_{flexible}$ shares the same distribution irrespectively to the current shape of any of them. The latter is an apparent illumination of the idea of lost identity. To compare, since the properties of the every time-independent Hamiltonian straightforwardly depend on the particularities of its shape and symmetries, the shape and symmetries appears as characteristics of the encapsulated identity.

Further in Chapter 5 [1] has been proven that the velocity of transitions from one binding region to another is always bounded. In result, the conjecture of adaptation meets the major general requirement to any process: that about the boundedness of the velocity by which the substance is transmitted, i.e. the velocity ansatz.

The obtained insensitivity of $\hat{H}_{flexible}$ to the chemical identity of the entities prompts that it is the fundament on which the universal protocol of coherence is to be built. The coherence is a process supposed to synchronize the relaxation of different entities so that their further behavior to be coherent. To elucidate this statement let us point out that the lack of synchronization gives rise to uncorrelated splitting of the entities among $\hat{H}_{flexible}$ and $\hat{H}_{rigid}$. In turn, those that belong to the latter sustain the amplification of the local fluctuations. On the other hand, the sensitivity of $\hat{H}_{flexible}$ to the spatio-temporal configuration of the local fluctuations renders the "flexible" parts of different entities non-identical even when their "rigid" parts ($\hat{H}_{rigid}$) are identical. Therefore, a successful coherence mechanism acts towards evening of current states through making all $\hat{H}_{flexible}$ identical. This task apparently calls for a non-local feedback that couples non-perturbatively different entities and operates as follows: the collision energy of the entities that are in states of $\hat{H}_{flexible}$ dissipates through excitation of local gapless modes, e.g. acoustic phonons in condensed matter; in turn they participate to $\hat{H}_{flexible}$ so that to induce a new transition. The latter dissipates non-radiatively again trough excitation of other local modes and so on. The process stops on making all $\hat{H}_{flexible}$ identical and synchronizing the further relaxation of all entities. The role of gapless modes is in providing an accelerating spread of the synchronization throughout the entire system. So, the non-local coupling based on the above feedback acts effectively as physical agent that renders each entity to "feel" distant ones, i.e. it behaves as analog to long-range interactions. The functional properties of the coherence have been studied in Chapter 6 of [1].

Let us now focus our attention how the coherence affects the self-organization. This is the task of the next section.

## 3. Feigenbaum Cascade, Fluctuation-Assisted Bifurcations and "Breathing" Structures

The major dilemma encountered now is whether the coherence saves the emergent hypothesis, i.e. whether the emergent properties on macroscopic scale are insensitive to the details of microscopic dynamics. A superficial look prompts that the dilemma is solved affirmatively for the emergent hypothesis because it seems that the coherence "averages out" the details of the local microscopic dynamics. In result, it is not to be anticipated that the coherence introduce anything new to the emergent hypothesis. However, the things are not as trivial as that. Let us first point out that first of all the "averaging out" of the local dynamics makes the reaction rate insensitive to the particularities of the reaction mechanism. However, this interferes with the core of the idea of the self-organization: the emergent properties are insensitive to the microscopic dynamics *though* the reaction rates in the corresponding reaction-diffusion equations are set on the particularities of the reaction collisions and the chemical identity.

Luckily, a closer look on the feedback that couples distant fluctuations shows that it operates in radically different from "averaging out" way. In Chapter 6 of [1] we have rigorously proved that the feedback selects that local dynamics which initially is in the most favorable local configuration and makes all other local dynamics equal to it. In other words, the feedback provides coherence by means of "imposing" one local dynamics onto the entire system. Note that though it varies from one location to another, the local dynamics preserves the identity of the reaction in a collision and thus the value of corresponding reaction rate is set on the particularities of the reaction mechanism and the chemical identity.

But what makes the above mechanism more advantageous than averaging out? Recalling that the latter is unstable to the amplification of the local fluctuations process, the coherence is justified by two crucial for its stability properties: as proven in Chapter 6 [1] it happens with accelerating velocity and irrespectively to the configurational morphology of the local fluctuations. It is obvious that the advantage of these properties is that they make the coherence stable to the local amplifications that appear later.

Outlining, the coherence is a stable process that saves the identity of reaction by making the global reaction rate equal to a particular local one irrespectively to the configurational morphology.

Now we come to the next fundamental problem: does the coherence provide separation of the time scales? The answer is immediate and it is positive: the coherence is temporary process that makes the intensive state variable, e.g. concentration, globally well-defined. Note that on starting the coherence, the intensive state variables are defined only locally due to the configurational disorder induced by interplay between the amplification of the local fluctuations and the lack of correlations among them. Therefore, the coherence provides the necessary ingredients for describing temporal evolution of a system in terms of state variables: the state variables and the reaction rates are globally well-defined.

Yet, the transition from microscopic dynamics to macroscopic evolution is not trivial. The key point is that the temporal behavior on microscopic scale is a discrete in time process: it starts with a coherence session followed by session of relaxation through $\hat{H}_{rigid}$ then again comes a coherence session and so on. Further we assume that the discreteness in time is scaling invariant which implies that there is no physical process that gives rise to correlations among time scales. This makes the corresponding evolutionary equations discrete mappings rather than differential equations:

$$\frac{\Delta \vec{x}}{\Delta t_i} = \hat{A}_i(\vec{x}) - \hat{R}_i(\vec{x}) + \hat{D}_i \bullet DivGrad\vec{x} \qquad (29)$$

where $\vec{x}$ is the vector of the state variables; $\Delta t_i$ is the duration of the $i-th$ session; $\hat{A}(\vec{x})$ and $\hat{R}(\vec{x})$ are the reaction rates; $D_i$ is the diffusion coefficient; the index $i$ serves to stress that the reaction rates vary from one session to another. Thus, the coherence renders replacement of the evolutionary equations with discrete mappings instead of differential equations. In turn, the discrete mappings give rise to novel properties such as Feigenbaum cascade. The advantage of our approach is that it puts the use of discrete mappings as evolutionary equations on mathematically rigorous and physical credible fundament.

The stochastisity of the reaction rates is another novel property introduced by the considered in &3 mechanism of coherence. But what new does it bring about? Recalling that the feedback picks up that local dynamics which is in the most favorable local configuration and makes all other equal to it, it becomes obvious that the configurational disorder renders the local dynamics selected in different sessions non-correlated. In result, the corresponding reaction rates fluctuate significantly with the coherence sessions. On the other hand, being product of one particular local dynamics, the global reaction rate depends straightforwardly on the microscopic dynamics. Luckily, it does not violate the core of the emergent hypothesis but essentially modifies it. Our next task it to show the details of that modification.

In Chapter 6 [1] it has been proven that the amplitude of the reaction rates remains permanently bounded. In turn, this makes possible to present (29) in the following form:

$$\frac{\Delta \vec{x}}{\Delta t_i} = \hat{A}_{av}(\vec{x}) + \hat{\eta}_{ai}(\vec{x}) - \hat{R}_{av}(\vec{x}) - \hat{\eta}_{ri}(\vec{x}) + \hat{D}_{av} \bullet DivGrad\vec{x} + \hat{d}_i \bullet DivGrad\vec{x} \qquad (30)$$

where and $\hat{R}_{av}(\vec{x})$ are the rate averages; $\hat{\eta}_{ai}(\vec{x})$, $\hat{\eta}_{ri}(\vec{x})$ and $\hat{d}_i$ are stochastic terms that come from the difference between the current and averaged rates; the index $i$ is put to stress on the stochastic nature of the corresponding terms. It should be stressed that, as proven in Chapter 1 [1], the additive decomposition to average and stochastic part is available at arbitrary statistics of the fluctuations only for bounded irregular series. Now, note that being averaged over the set of all admissible local configurations, $\hat{A}_{av}(\vec{x})$, $\hat{R}_{av}(\vec{x})$ and $\hat{D}_{av}$ are insensitive to the microscopic dynamics. The advantage of this insensitivity is best utilized by means of the following consideration: the stochastic terms in (30) make its solution irregular function of the spatio-temporal variables that fluctuates around the solution of the following equation:

$$\frac{\Delta \vec{x}_{det}}{\Delta t_i} = \hat{A}_{av}(\vec{x}_{det}) - \hat{R}_{av}(\vec{x}_{det}) + \hat{D}_{av} \bullet DivGrad\vec{x}_{det} \qquad (31)$$

Obviously, (31) coincides with the familiar deterministic reaction-diffusion equations whose reaction rates are averaged out over the microscopic dynamics. Eq.(31) along with the boundary conditions imposed on the system gives rise to macroscopic structures, traveling waves etc. whose details depend on the non-linearities involved and the particularities of the boundaries. Though the properties of those structures are non-generic, once emerged, they stay for ever and never exhibit fluctuations. Thus, they can serve as metaphorical "skeleton" around which the solution of (30) permanently fluctuates. Until those fluctuations are small, they can be considered figuratively as "breathing" of the "skeleton". However, due time course the size of a fluctuation can become as large as the distance to a bifurcation point; in result, this gives rise to new type dynamical transitions, called hereafter fluctuation-assisted bifurcations. They have two major properties: (i) the always have temporary effect – any of them appears when the size of a fluctuation happens to cover the distance to the bifurcation point and lasts until that amplitude of the fluctuation is large enough to keep the system beyond the bifurcation point; (ii) the fluctuation-assisted bifurcations do not need sharp setting of the control parameters to their bifurcation values; on the contrary, since the fluctuations can be considered as coming from an effective shift of the control parameters in (31), reaching the appropriate size of fluctuations is equivalent to adjusting the control parameters to their bifurcation values.

## Discussion

The credibility of every theory is verified when it is built on self-consistent basis. It implies that the aspects of the theory associated with different postulates must be in accordance with each other. However, as it has been proven in &1, the theory of self-organization suffers serious flaw: it lets the velocity of transmitting substance to be arbitrary. Furthermore, the situation becomes even more difficult because the interplay between the velocity ansatz and the short-range interactions produces amplification of the local fluctuations that in turn sustains the configurational disorder. The route to avoiding an inevitable breakdown goes through reconsidering the idea of interaction. As a result, the dynamics of the fluctuations is approached in a radically novel way grounded on the interplay between the velocity ansatz and the idea of coherence.

Summarizing, the developed by us novel approach to stochastic dynamics not only saves the idea of self-organization but extends the set of emergent properties including to them phenomena like Feigenbaum cascade and fluctuation-assisted bifurcations. Besides, it makes the emergent structures alive: they breathe!